\documentclass[prb,amsmath,superscriptaddress,twocolumn,showpacs]{revtex4-1}
\usepackage{amssymb,amsmath}   % need for subequations
\usepackage[dvips]{graphicx}   % need for figures
\usepackage{color}      % use if color is used in text
\usepackage{subfigure}  % use for side-by-side figures
\usepackage{hyperref}   % use for hypertext links, including those to external documents and URLs
\usepackage{gensymb}
\usepackage{epstopdf}
\usepackage{natbib}
\usepackage{enumerate}
\usepackage{mathbbol}
\usepackage{braket}
\usepackage{soul}
\usepackage{ulem}
\usepackage[makeroom]{cancel}

\DeclareMathOperator{\Tr}{Tr}

\begin{document}

\title{Spin filtering in all-electrical three-terminal interferometers}
\author{S. Matityahu} \email{matityas@post.bgu.ac.il}
\affiliation{Department of Physics, Ben-Gurion University, Beer Sheva 84105, Israel}
\affiliation{Department of Physics, NRCN, P.O. Box 9001, Beer-Sheva 84190, Israel}
%
%\author{Yasuhiro Utsumi}
%\affiliation{Department of Physics Engineering, Faculty of
%Engineering, Mie University, Tsu, Mie, 514-8507, Japan}
%
\author{A. Aharony}
\affiliation{Department of Physics, Ben-Gurion University, Beer Sheva 84105, Israel}
%\affiliation{Ilse Katz Center for Meso- and Nano-Scale Science and Technology, Ben-Gurion University, Beer Sheva 84105, Israel}
\affiliation{Raymond and Beverly Sackler School of Physics and Astronomy, Tel Aviv University, Tel Aviv 69978, Israel}
\author{O. Entin-Wohlman}
\affiliation{Department of Physics, Ben-Gurion University, Beer Sheva 84105, Israel}
%\affiliation{Ilse Katz Center for Meso- and Nano-Scale Science and Technology, Ben-Gurion University, Beer Sheva 84105, Israel}
\affiliation{Raymond and Beverly Sackler School of Physics and Astronomy, Tel Aviv University, Tel Aviv 69978, Israel}
\author{C. A. Balseiro}
\affiliation{Centro At\'{o}mico Bariloche and Instituto Balseiro, Comisi\'{o}n Nacional de Energ\'{i}a At\'{o}mica, 8400 S. C. de Bariloche, Argentina}
\affiliation{Consejo Nacional de Investigaciones Cient\'{i}ficas y T\'{e}cnicas (CONICET), Argentina}
\date{\today}

\begin{abstract}
Unlike the two-terminal device, in which the time-reversal invariant spin-orbit interaction alone cannot polarize the spins, such a polarization can be generated when electrons from one source reservoir flow into two (or more) separate drain reservoirs. We present analytical solutions for two examples. First, we demonstrate that the electrons transmitted through a ``diamond" interferometer into two drains can be simultaneously fully spin-polarized along different tunable directions, even when the two arms of the interferometer are not identical. Second, we show that a single helical molecule attached to more than one drain can induce a significant spin polarization in electrons passing through it. The average polarization remains non-zero even when the electrons outgoing into separate leads are eventually mixed incoherently into one absorbing reservoir. This may explain recent experiments on spin selectivity of  certain helical-chiral  molecules.
\end{abstract}

\pacs{}

\maketitle
\section{Introduction} \label{Sec1}
One of the major aims of spintronics is the generation and manipulation of spin-polarized electrons in semiconductors.~\cite{WSA01,ZI04} The coherent control of the electron spin~\cite{HR08} has important implications for future spintronic devices,~\cite{DDA07} as well as for spin-based quantum computation.~\cite{Nielsen&Chuang,LD98} Two classes of spin-based architectures may be distinguished. The first uses  static qubits, i.e., electrons localized  in spatially-confined systems, such as quantum dots.~\cite{LD98} In this type of architecture, single- and two-qubit gates function by a suitable time-dependent tuning of magnetic and electric fields. The second class is based on flying qubits, i.e.,  mobile electrons which move through the circuit, passing via quantum gates implemented in predefined areas by static electric and magnetic fields.~\cite{PAE04} In this paper we consider the second class of spin qubits and focus on spin filters: devices which polarize the spins of electrons going through them along tunable directions, or equivalently, write quantum information on these mobile qubits. Specifically, we concentrate on time-reversal symmetric devices, operating in the absence of external magnetic fields. We first discuss mesoscopic interferometers, which can induce full polarization of the outgoing electron spins, along tunable directions, and thus can serve as spin filters. We then consider helical molecules, which typically lead to only partial polarization (sometimes called spin selectivity).

A natural source of spin polarization is the spin-orbit interaction (SOI).~\cite{Dresselhaus,Rashba,Winkler} This interaction generates a momentum-dependent effective magnetic field which operates on an electron moving in an electric field. This effective field  couples to the electronic spin, which then  precesses around it. When electrons move on a single one-dimensional (1D) wire, all their spins rotate by the same amount, hence no net polarization appears. Technically, the SOI can be removed by a gauge transformation.~\cite{MH92} The effect of the SOI becomes nontrivial in transport through more than one electronic path, for instance by allowing for quantum interference.~\cite{BD15} Indeed, several groups proposed spin filters based on a single loop connected to two terminals by single-channel 1D leads.~\cite{MB04,CR06,HN07,AA11,MS13a} If time-reversal symmetry (TRS) is conserved in such networks and the scattering matrix of the device is unitary (meaning that the number of particles scattered at a certain energy is conserved), then the $2\times2$ transmission and reflection matrices (in spin space) must have degenerate eigenvalues.~\cite{BCWJ97,BJH08} As a result, all directions of the spin polarization are equally probable, and there cannot be any spin filtering. Since the SOI conserves TRS, a spin filter based on a two-terminal device requires departure from the above assumptions. One way to achieve this is by breaking TRS, e.g.\ by adding a magnetic field.~\cite{MB04,CR06,HN07,AA11,MS13a}

\begin{figure}[ht]
\centering
\includegraphics[width=0.3\textwidth,height=0.25\textheight]{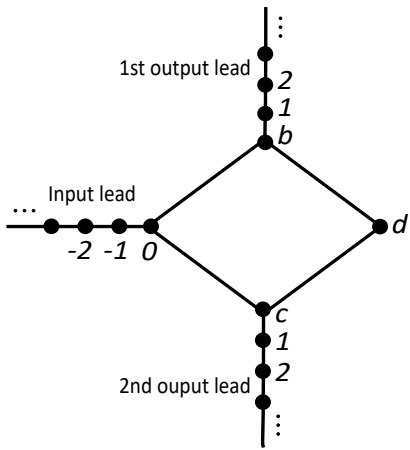}
\caption{\label{fig:three terminal interferometer} Illustration of a three-terminal diamond interferometer. The electrons moving on the diamond edges (of length $L$) are subjected to spin-orbit interactions.}
\end{figure}

Alternatively, one may increase the number of terminals connected to the device and thus generate a finite spin polarization without exploiting ferromagnetic electrodes and without applying magnetic fields. All-electrical single-loop spin filters based on three-terminal devices have been studied before;~\cite{KAA01,PTP04,YM05,FP06,YT12} in particular, F\"{o}ldi {\it et al.}~\cite{FP06}
demonstrated that a symmetric-ring interferometer attached to one source and two drain terminals can act as a spin beam-splitter, which polarizes the electrons in the output leads along tunable directions. In the first part of this paper we extend this result to a more realistic setup. By analyzing the diamond interferometer shown in Fig.~\ref{fig:three terminal interferometer} we find that it can serve as a perfect spin beam-splitter even when its two arms are not completely identical. We obtain the spin-filtering conditions for the two output leads and show that the two outgoing electron beams can be fully polarized simultaneously along different directions, which are determined solely by the parameters of the SOI.

In spin filters based on the single-loop interferometers discussed above, the outgoing electrons are separated into two beams; the quantity of interest is then the spin polarization of each of these beams. However, other configurations, in which the output leads are connected to a single reservoir where the transmitted electrons are mixed incoherently, are also possible. Then the relevant spin polarization is  the total (average) polarization of all the transmitted electrons. As we argue below, this may be the case in the chiral-induced spin selectivity (CISS) effect. This remarkable effect has revealed the ability of chiral organic molecules to act as efficient spin filters,~\cite{RSG06,GB11,XZ11,MD13,EH15,KM15,NR15}  thus opening the route for organic spintronic applications.~\cite{NWJM07,DVA,BC13} This is achieved  without ferromagnetic electrodes and in the absence of magnetic fields: the organic molecules serve as active elements rather than passive ones.~\cite{BDO13} It was found experimentally that the conductance of a chiral molecule adsorbed on a ferromagnetic substrate and connected to a metallic nanoparticle at its other edge,\cite{XZ11,KM15} depends on the polarization direction of the substrate. Other experiments detected the spin polarization of photoelectrons transmitted through a self-assembled monolayer of molecules adsorbed on a metallic substrate.~\cite{GB11,MD13,KM15} Here we model the experimental setup by a multi-terminal helical configuration, and show that the total polarization of the outgoing electrons can be quite significant.

The experimental observations of the CISS effect were followed by numerous theoretical proposals aiming to explain them.~\cite{GR12,GR13,GAM12,GAM14a,VS14,ME12,EAA13,GJ13,MS16,WHN16} However, there is still no consensus on the appropriate theory. To explain the CISS effect without invoking a magnetic field, several earlier theories calculated the current between a single source and a single drain, but effectively added coupling of the system to additional reservoirs. References~\onlinecite{GAM12,GAM14a,WHN16} used a phenomenological method due to B\"{u}ttiker,~\cite{Buttiker} in which phase-breaking (or dephasing) processes are modeled by additional electron reservoirs coupled to the system via fictitious voltage probes, subjected to the condition of zero net current. Some of the electrons enter the fictitious probes and have their phases randomized before they are re-injected into the system. Alternatively, Ref.~~\onlinecite{MS16} broke the unitarity of the two-terminal scattering matrix by adding leakage of electrons into additional reservoirs.\cite{AA02,MS13b} As we argue below, the results of these calculations apply only when the additional (leakage) leads are held at the same chemical potential as the output lead. Unlike Ref.~\onlinecite{MS16}, which calculated only the spin polarization in a single output lead, we calculate below the polarization in all the output leads and also their total polarization. We find that a significant spin polarization can be achieved in the fully coherent and unitary transport regime, by considering a multi-terminal configuration in which electrons are allowed to exit the molecule into more than one terminal.

A similar model was used to study the effects of various contacts connected to a double-helical molecule.~\cite{GAM14b} As expected, the net polarization vanished in all the examined two-terminal configurations. In contrast, a finite spin polarization was found when the double-helical molecule was connected in a multi-terminal configuration. The polarization was negligible when the two strands are identical, and approached to zero in the limit of a single 1D strand. We study a minimal (simpler) model for a single helical molecule in a three-terminal configuration and show that a significant polarization can be obtained, provided that interference between different electronic paths is taken into account. The model also yields that the sign of the polarization is reversed upon reversing the chirality sense of the molecule, as observed in experiments~\cite{NR15}.

The outline of the paper is as follows: in Sec.~\ref{Sec2} we present the tight-binding model used to study spin-dependent transport in a finite quantum network with SOI, which is connected to an arbitrary number of reservoirs by tight-binding chains (``leads"), and express the spin polarization of the electrons in each lead in terms of the scattering matrix of the network. We then solve for the transmission of a three-terminal diamond interferometer, and find the conditions for full spin filtering in each output lead (Sec.~\ref{Sec3}). In Sec.~\ref{Sec4} we apply the multi-terminal model to a single helical molecule, with further-neighbor hopping which accounts for interference, and calculate the total spin polarization of the outgoing electrons along the symmetry axis of the molecule. The results are discussed and summarized in Sec.~\ref{Summary}.
%%%%%%%%%%%%%%%%%%%%%%%%%%%%%%%%%%%%%%%%%%%%%%%%%%%%%%%%%%%%
\section{General considerations}
\label{Sec2}
To study spin-dependent electron transport in the presence of SOI, we exploit the tight-binding formalism in which electrons hop between discrete sites. In this description, the
Schr\"{o}dinger equation for the spinor $\ket{\psi^{}_{\beta}}$ at site $\beta$ is
\begin{align}
\label{eq:Schrodinger}&\left(E-\varepsilon^{}_{\beta}\right)\ket{\psi^{}_{\beta}}=-\sum_{\alpha}J_{\alpha\beta}U_{\alpha\beta}\ket{\psi^{}_{\alpha}}\!\ ,
\end{align}
where $E$ is the electron energy, $\varepsilon^{}_{\beta}$ is the on-site energy, $J_{\alpha\beta}$ is the hopping amplitude from site $\alpha$ to site $\beta$ (which can be chosen to be real) and $U_{\alpha\beta}$ is a $2\times 2$ unitary matrix which describes the spin precession of an electron moving from site $\alpha$ to site $\beta$. Generally, these unitary matrices are of the form~\cite{OY92}
\begin{align}
\label{eq:SOI matrices}&U_{\alpha\beta}=\exp[i\mathbf{K}_{\alpha\beta}\cdot\boldsymbol\sigma]\ ,
\end{align}
with $\boldsymbol\sigma$ being the vector of Pauli matrices. The vector $\mathbf{K}_{\alpha\beta}$ depends on the specific type of SOI.~\cite{AA11,MS13a} Its magnitude scales with the strength of the local SOI and its direction depends on the type of SOI and on the direction of the bond $\alpha\beta$. Specifically, we assume that the semi-infinite leads (see Fig. \ref{fig:three terminal interferometer}) are free of the SOI; the nearest-neighbor hopping amplitude on those leads (the same for all bonds) is denoted $J_{0}$, and the on-site energies there are set to zero.

In the context of mesoscopic interferometers patterned from a two-dimensional electron gas confined at the interface of narrow-gap semiconductor heterostructures, the relevant types of SOI are the Dresselhaus~\cite{Dresselhaus} and the Rashba~\cite{Rashba} interactions. The Rashba SOI originates in broken structural inversion symmetry along the growth direction of the quantum well (taken to be  $\hat{\bold{z}})$, and its magnitude can be tuned by applying an electric field along that axis, via a gate voltage.~\cite{NJ97,HJP98,GD00,KT02,KM06,BT06,LD12} The Dresselhaus SOI results from broken bulk inversion symmetry, and its magnitude is nearly independent of the applied electric field. For these two interactions, the vectors $\mathbf{K}_{\alpha\beta}$   are~\cite{AA11,MS13a}
\begin{align}
\label{eq:K vectors 1}&\mathbf{K}_{\alpha\beta}=\theta_{D}(-\hat{d}_{\alpha\beta,x},\hat{d}_{\alpha\beta,y},0)+\theta_{R}(-\hat{d}_{\alpha\beta,y},\hat{d}_{\alpha\beta,x},0)\ ,
\end{align}
where $\theta_{D}$ and $\theta_{R}$ are the spin precession angles due to Dresselhaus and Rashba SOI, respectively. These angles  are proportional to the strength of the corresponding SOI mechanism and to the length of the bond $\alpha\beta$, whose direction is along the unit vector $\hat{\mathbf{d}}_{\alpha\beta}=(\hat{d}_{\alpha\beta,x},\hat{d}_{\alpha\beta,y},0)$. For chiral molecules, a more plausible assumption about the specific SOI is that it is due to the local electric fields within the molecule. In this case~\cite{OY92}
\begin{align}
\label{eq:K vectors 2}&\mathbf{K}_{\alpha\beta}=\lambda\mathbf{d}_{\alpha\beta}\times\mathbf{E}_{\alpha\beta}\ ,
\end{align}
where $\lambda$ is the parameter representing the SOI strength, $\mathbf{d}_{\alpha\beta}$ is the vector along the bond $\alpha\beta$, and $\mathbf{E}_{\alpha\beta}$ is the average electric field acting on an electron along this bond. Below we assume that for a helical molecule this field points in the radial direction, due to the potential which confines the electrons to move on the helical cylinder. In atoms $\lambda=e/(4m_{e}c^{2})$, where $e$ and $m_{e}$ are the free electron charge and mass, respectively, and $c$ is the speed of light. The strength of the SOI is enhanced in curved structures such as carbon nanotubes and chiral helical molecules, as compared to its value in flat configurations.~\cite{NR15,HHD06,KF08,JJS09,CHOH09,JSH10,JTS11,SGA13,LEA15}

Quite generally, the spin-resolved currents in the leads can be expressed using the Landauer-B\"{u}ttiker formalism, which yields the charge and the spin currents in terms of the electronic populations in the various reservoirs, and the transmissions between the leads. When the latter are free of SOI's and magnetic fields and the electrons' reservoirs to which they are attached are not polarized, the charge and spin currents in lead $i$ read~\cite{SM07}
\begin{align}
\label{eq:Charge and spin}&I^{(C)}_{i}=e\int\frac{dE}{2\pi\hbar}\sum_{j\neq i}\left[f_{i}(E)-f_{j}(E)\right]T^{(C)}_{i,j}(E)\ ,\nonumber\\
&I^{(S)}_{i}=\int\frac{dE}{4\pi\hbar}\sum_{j\neq i}\left[f_{i}(E)-f_{j}(E)\right]T^{(S)}_{i,j}(E)\ ,
\end{align}
where $f_{i}(E)=\left(\exp[(E-\mu_{i})/k_{B}T]+1\right)^{-1}$ is the Fermi-Dirac distribution  of the $i$th reservoir, whose chemical potential is $\mu_{i}$. In Eq.~(\ref{eq:Charge and spin}),
\begin{align}
T^{(C)}_{i,j}(E)&=\sum_{\sigma,\sigma'}T_{i\sigma,j\sigma'}(E)\ ,\nonumber\\
T^{(S)}_{i,j}(E)&=\sum_{\sigma,\sigma'}\sigma T_{i\sigma,j\sigma'}(E)\ ,
\end{align}
are the charge and spin transmissions between lead $j$ and lead $i$. These are derived within scattering theory, by calculating the scattering matrix  $\mathcal{S}(E)$ (whose elements are $i\sigma, j\sigma'$) of the  setup at energy $E$; the various transmissions are then $T_{i\sigma,j\sigma'}(E)=|\mathcal{S}_{i\sigma,j\sigma'}(E)|^{2}$. The indices $\sigma$ and $\sigma'$ are the eigenvalues of the spins along an arbitrary quantization axis.

When all the leads but one are attached to reservoirs with the same chemical potential, lower than that of the one coupled to the remaining lead (say lead 0), the unitarity of the scattering matrix implies that all the currents can be expressed in terms of the reflection and transmissions from the source lead 0; in this case only a single column (containing $2\times 2$ matrices in spin space) of the scattering matrix is needed. This is the situation implicitly assumed in our previous paper,~\cite{MS16} as well as in other  works;~\cite{YM05,FP06,GAM14b}  the same configuration is exploited below. Presumably, this configuration applies to the transport experiments carried  out on chiral organic molecules, in which a single voltage source is applied to the two edges of the molecule.~\cite{XZ11}

To calculate the transmissions $T_{i\sigma,j\sigma'}(E)$'s we assume a scattering state at a given energy $E$, in which the site spinors on the leads are given by
\begin{widetext}
\begin{align}
\label{eq:scattering state 3ter}&\ket{\psi^{}_{n}}=\begin{cases} \ket{\chi^{}_{in}} e^{ik_{0}(n-n_{in})}+r\ket{\chi^{}_{r}} e^{-ik_{0}(n-n_{in})} & \text{input lead\ ,}\\
t^{(1)}\ket{\chi^{(1)}_{t}}e^{ik_{0}(n-n_{out,1})} & \text{1st output lead\ ,}\\
t^{(2)}\ket{\chi^{(2)}_{t}}e^{ik_{0}(n-n_{out,2})} & \text{2nd output lead\ , {\it etc}}\ .
\end{cases}
\end{align}
\end{widetext}
Here $\ket{\chi^{}_{in}}$, $\ket{\chi^{}_{r}}$ and $\ket{\chi^{(n)}_{t}}$ ($n=1, 2, ...$) are the  incoming, reflected, and transmitted spinors, respectively, and the corresponding reflection and transmission amplitudes are $r$ and $t^{(n)}$. The total incoming current is normalized to a unit of particle flux. The wave vector $k_{0}$ (in units of the inverse of the lattice constant of the leads~\cite{Comment1}) obeys the dispersion relation $E=-2J_{0}\cos k_{0}$; the indices $n_{in}$, $n_{out,1}$, and $n_{out,2}$, {\it etc.}, stand for the sites connecting the system to the leads (e.g.,  $n_{in}=0,~n_{out,1}=n_{b}=0,~n_{out,2}=n_{c}=0$ for the interferometer in Fig.~\ref{fig:three terminal interferometer}).

The reflection and transmission amplitude matrices, $\mathcal{R}$ and $\mathcal{T}_{n}$, defined by
\begin{align}
\label{eq:R and T matrices 1}&r\ket{\chi^{}_{r}}=\mathcal{R}\ket{\chi^{}_{in}}\!\ ,\nonumber\\
&t^{(n)}\ket{\chi^{(n)}_{t}}=\mathcal{T}_{n}\ket{\chi^{}_{in}}\!\ ,
\end{align}
are obtained by inserting the scattering wave
Eq.~(\ref{eq:scattering state 3ter}) into the Schr\"{o}dinger Eq.~(\ref{eq:Schrodinger}).
As we find in the examples studied below, one can write these matrices in the form
\begin{align}
\label{eq:R and T matrices 2}&\mathcal{R}=r\mathbf{1},\nonumber\\
&\mathcal{T}_{n}=t^{(n)}_{+}\ket{\hat{\bold{n}}_{n}}\bra{\hat{\bold{n}}}+t^{(n)}_{-}\ket{-\hat{\bold{n}}_{n}}\bra{-\hat{\bold{n}}}\ ,
\end{align}
where $\mathbf{1}$ is the $2\times 2$ unit matrix and the spinors $\ket{\pm\hat{\bold{n}}}$ and
$\ket{\pm\hat{\bold{n}}_n}$ are spin eigenstates with the polarization along $\pm\hat{\bold{n}}$ and $\pm\hat{\bold{n}}_n$ (see below); they depend on the geometry of the system and on the specific SOI. Note that the scalar form of the reflection amplitude matrix follows from the self-duality of the scattering matrix.~\cite{BCWJ97} Due to TRS, the reflected electrons are never polarized when there is a single ``input" channel, for any number of ``output" leads.\cite{BJH08}

For an unpolarized incident beam, the polarization of the outgoing beam in the $n$th output lead along $\hat{\bold{n}}_{n}$ is
\begin{align}
\label{eq:polarization1}&P^{(n)}_{\hat{\bold{n}}_{n}}\equiv\frac{\Tr\left[\mathcal{T}^{\dag}_{n}\left(\hat{\bold{n}}_{n}\cdot\boldsymbol\sigma\right)\mathcal{T}_{n}\right]}{\Tr\left[\mathcal{T}^{\dag}_{n}\mathcal{T}_{n}\right]}=\frac{|t^{(n)}_{+}|^{2}-|t^{(n)}_{-}|^{2}}{|t^{(n)}_{+}|^{2}+|t^{(n)}_{-}|^{2}}\ .
\end{align}
The total polarization along an arbitrary unit vector $\hat{\bold{z}}$ is given by the weighted average of the polarizations in each lead with respect to the total transmission into each lead,
\begin{align}
\label{eq:polarization2}&P_{\hat{\bold{z}}}\equiv \sum_n q_{n}P^{(n)}_{\hat{\bold{n}}_{n}}\hat{\bold{n}}_{n}\cdot\hat{\bold{z}}\ ,
\end{align}
where
\begin{align}
\label{eq:fractional transmission}&q_{n}=\frac{\Tr\left[\mathcal{T}^{\dag}_{n}\mathcal{T}_{n}\right]}{\sum_{n'}
\Tr\left[\mathcal{T}^{\dag}_{n'}\mathcal{T}^{}_{n'}\right]}\ .
\end{align}
These expressions are used in the following to analyze specific setups.
%%%%%%%%%%%%%%%%%%%%%%%%%%%%%%%%%%%%%%%%%%%%%%%%%%%%%%%%%%%%
\section{Three-terminal diamond interferometer} \label{Sec3}
Applying the general formulation to the three-terminal diamond interferometer shown in Fig.~\ref{fig:three terminal interferometer}, the Schr\"{o}dinger equations for the spinors at sites $0$, $b$, $c$,  and $d$ are
\begin{align}
\label{eq:tight binding 3ter 1}
&\left(E-\varepsilon^{}_{0}\right)\ket{\psi^{}_{0}}=-J_{0b}U^{\dag}_{0b}\ket{\psi^{}_{b}}-J_{0c}U^{\dag}_{0c}\ket{\psi^{}_{c}}-J_{0}\ket{\psi^{(0)}_{-1}}\!\ ,\nonumber\\
&\left(E-\varepsilon^{}_{b}\right)\ket{\psi^{}_{b}}=-J_{0b}U_{0b}\ket{\psi^{}_{0}}-J_{bd}U^{\dag}_{bd}\ket{\psi^{}_{d}}-J_{0}\ket{\psi^{(1)}_{1}}\!\ ,\nonumber\\
&\left(E-\varepsilon^{}_{c}\right)\ket{\psi^{}_{c}}=-J_{0c}U_{0c}\ket{\psi^{}_{0}}-J_{cd}U^{\dag}_{cd}\ket{\psi^{}_{d}}-J_{0}\ket{\psi^{(2)}_{1}}\!\ ,\nonumber\\
&\left(E-\varepsilon^{}_{d}\right)\ket{\psi^{}_{d}}=-J_{bd}U_{bd}\ket{\psi^{}_{b}}-J_{cd}U_{cd}\ket{\psi^{}_{c}}\!\ .
\end{align}
The spinors with a single subscript stand for the wave functions at the four sites $0$, $b$, $c$, and $d$ which define the interferometer. Spinors with superscripts represent the sites along the leads, where the superscript specifies the lead (0 for the input lead and 1, 2 for the two output leads) and the subscript specifies the site within that lead (Fig.~\ref{fig:three terminal interferometer}). Eliminating $\ket{\psi^{}_{d}}$ from Eqs.~(\ref{eq:tight binding 3ter 1}), one finds
\begin{align}
\label{eq:tight binding 3ter 2}
\left(E-\varepsilon^{}_{0}\right)\ket{\psi^{}_{0}}=&-J_{0b}U^{\dag}_{0b}\ket{\psi^{}_{b}}-J_{0c}U^{\dag}_{0c}\ket{\psi^{}_{c}}-J_{0}\ket{\psi^{(0)}_{-1}}\!\ ,\nonumber\\
\left(E-y_{b}\right)\ket{\psi^{}_{b}}=&-J_{0b}U_{0b}\ket{\psi^{}_{0}}+JU^{\dag}_{bd}U_{cd}\ket{\psi^{}_{c}}-J_{0}\ket{\psi^{(1)}_{1}}\!\ ,\nonumber\\
\left(E-y_{c}\right)\ket{\psi^{}_{c}}=&-J_{0c}U_{0c}\ket{\psi^{}_{0}}+JU^{\dag}_{cd}U_{bd}\ket{\psi^{}_{b}}-J_{0}\ket{\psi^{(2)}_{1}}\!\ ,
\end{align}
where
\begin{align}
\label{eq:3ter parameters 1}
y_{\alpha}=\varepsilon^{}_{\alpha}+\frac{J^{2}_{\alpha d}}{E-\varepsilon^{}_{d}} \quad (\alpha=b,c)\ ,
%\nonumber\\
\ \ \ \ J=\frac{J_{bd}J_{cd}}{E-\varepsilon^{}_{d}}\ .
\end{align}
Inserting the scattering state~(\ref{eq:scattering state 3ter}) into Eq.~(\ref{eq:tight binding 3ter 2}) and solving for $r\ket{\chi^{}_{r}}$, $t^{(1)}\ket{\chi^{(1)}_{t}}$, and $t^{(2)}\ket{\chi^{(2)}_{2}}$, gives the $2\times 2$ reflection and transmission amplitude matrices [see Eq. (\ref{eq:R and T matrices 2})]
\begin{align}
\label{eq:R and T matrices 3}\mathcal{R}=&-\mathbf{1}-2iJ_{0}Y\sin k_{0}\left[Z-J_{0b}J_{0c}J\left(u+u^{\dag}\right)\right]^{-1}\!\ ,\nonumber\\
\mathcal{T}_{1}=&\,\,2iJ_{0}\sin k_{0}\left[Z-J_{0b}J_{0c}J\left(u+u^{\dag}\right)\right]^{-1}U_{0b}\nonumber\\
&\times\left(JJ_{0c}u-X_{c}J_{0b}\right)\!\ ,\nonumber\\
\mathcal{T}_{2}=&\,\,2iJ_{0}\sin k_{0}\left[Z-J_{0b}J_{0c}J\left(u+u^{\dag}\right)\right]^{-1}U_{0c}\nonumber\\
&\times\left(JJ_{0b}u^{\dagger}-X_{b}J_{0c}\right)\!\ ,
\end{align}
with
\begin{align}
\label{eq:3ter parameters 2}&X_{0}=\varepsilon^{}_{0}+J_{0}e^{-ik_{0}}\ ,\nonumber\\
&X_{\alpha}=y_{\alpha}+J_{0}e^{-ik_{0}} \quad (\alpha=b,c)\ ,\nonumber\\
&Y=J^{2}-X_{b}X_{c}\ ,\nonumber\\
&Z=X_{0}Y+X_{c}J^{2}_{0b}+X_{b}J^{2}_{0c}\ .
\end{align}
The unitary matrix $u=U^{\dagger}_{0b}U^{\dagger}_{bd}U_{cd}U_{0c}$ in Eq.~(\ref{eq:R and T matrices 3}) represents anticlockwise hopping from site 0 back to site 0 around the loop. This matrix is of the general form
\begin{align}
\label{eq:u matrix}&u=\exp[i\omega\hat{\bold{n}}\cdot\boldsymbol{\sigma}]=\cos\omega+i\sin\omega\hat{\bold{n}}\cdot\boldsymbol{\sigma}\ ,
\end{align}
where $\omega$ is the phase accumulated around the loop due to the SOI-induced spin precession. The eigenspinors of $u$, $\ket{\pm\hat{\bold{n}}}$, are the eigenstates of the spin component $\hat{\bold{n}}\cdot\boldsymbol{\sigma}$ along the unit vector $\hat{\bold{n}}$; that is,
\begin{align}
u\ket{\pm\hat{\bold{n}}}=\exp[\pm i\omega]\ket{\pm\hat{\bold{n}}}\ .
\end{align}
The SOI phase $\omega$ and the direction $\hat{\bold{n}}$ are determined  by the SOI on the edges of the interferometer. As stated before Eq.~(\ref{eq:K vectors 1}), the angle $\theta_R$ due to the Rashba SOI can be controlled by a gate voltage, as demonstrated in several experiments.~\cite{NJ97,HJP98,GD00,KT02,KM06,BT06,LD12} The expressions for $\omega$ and $\hat{\bold{n}}$ in the presence of both the Rashba and the Dresselhaus SOIs were analyzed in Refs.~\onlinecite{AA11} and~\onlinecite{MS13a}.

As expected [see the first of Eqs.~(\ref{eq:R and T matrices 2})], the reflection amplitude matrix $\mathcal{R}$, given by the first of Eqs.~(\ref{eq:R and T matrices 3}) in conjunction with Eq.~(\ref{eq:u matrix}), is proportional to the unit matrix  with the reflection amplitude
\begin{align}
\label{eq:reflection amplitude}&r=-1-\frac{2iJ_{0}Y\sin k_{0}}{Z-2J_{0b}J_{0c}J\cos\omega}\ .
\end{align}
To find the transmission amplitudes, we expand the incoming spinor $\ket{\chi^{}_{in}}$ in the basis $\{\ket{\hat{\bold{n}}},\ket{-\hat{\bold{n}}}\}$,  $\ket{\chi^{}_{in}}=c^{}_{+}\ket{\hat{\bold{n}}}+c^{}_{-}\ket{-\hat{\bold{n}}}$. Then, for $\ket{\chi^{}_{in}}=\ket{\pm\hat{\bold{n}}}$, Eqs.~(\ref{eq:R and T matrices 3}) and~(\ref{eq:u matrix}) yield the spinors of the electrons transmitted into the output leads,
\begin{align}
\label{eq:3ter_transmitted}&t^{(1)}_{\pm}\ket{\chi^{(1)}_{t,\pm}}=\frac{2iJ_{0}\sin k_{0}\left(JJ_{0c}e^{\pm i\omega}-X_{c}J_{0b}\right)}{Z-2J_{0b}J_{0c}J\cos\omega}U_{0b}\ket{\pm\hat{\bold{n}}}\!\ ,\nonumber\\
&t^{(2)}_{\pm}\ket{\chi^{(2)}_{t,\pm}}=\frac{2iJ_{0}\sin k_{0}\left(JJ_{0b}e^{\mp i\omega}-X_{b}J_{0c}\right)}{Z-2J_{0b}J_{0c}J\cos\omega}U_{0c}\ket{\pm\hat{\bold{n}}}\!\ .
\end{align}
The transmission amplitude matrices $\mathcal{T}_{1}$ and $\mathcal{T}_{2}$ can therefore be written in the form of Eq.~(\ref{eq:R and T matrices 2}) with
\begin{align}
\label{eq:transmission eigenvalues}&t^{(1)}_{\pm}=\frac{2iJ_{0}\sin k_{0}\left(JJ_{0c}e^{\pm i\omega}-X_{c}J_{0b}\right)}{Z-2J_{0b}J_{0c}J\cos\omega}\ ,\nonumber\\
&t^{(2)}_{\pm}=\frac{2iJ_{0}\sin k_{0}\left(JJ_{0b}e^{\mp i\omega}-X_{b}J_{0c}\right)}{Z-2J_{0b}J_{0c}J\cos\omega}\ ,
\end{align}
and
\begin{align}
\label{eq:transmitted spinors}&\ket{\pm\hat{\bold{n}}_{1}}=U_{0b}\ket{\pm\hat{\bold{n}}}\!\ ,\nonumber\\
&\ket{\pm\hat{\bold{n}}_{2}}=U_{0c}\ket{\pm\hat{\bold{n}}}\!\ .
\end{align}
Note that the two terms in the brackets in Eqs.~(\ref{eq:transmission eigenvalues}) reflect the interference between the different arms of the interferometer. When one of the arms is blocked (e.g., by setting $J^{}_{0b}=0$), one obtains $|t^{(n)}_{+}|=|t^{(n)}_{-}|$ and the polarization~(\ref{eq:polarization1}) vanishes, as expected for a 1D wire.

The form~(\ref{eq:R and T matrices 2}) of the transmission amplitude matrices indicates that the electrons transmitted into  terminal 1 (2) are fully spin-polarized when one of the transmission eigenvalues $t^{(1)}_{\pm}$ ($t^{(2)}_{\pm}$) vanishes. For instance, if $t^{(1)}_{-}=0$ ($t^{(1)}_{+}=0$) the electrons transmitted into  terminal 1 are fully polarized along the direction $\hat{\bold{n}}_{1}$ ($-\hat{\bold{n}}_{1}$), with the amplitude $c^{}_{+}t^{(1)}_{+}$ ($c^{}_{-}t^{(1)}_{-}$). The outgoing currents in these leads thus contain information on the incoming spinor, via the factors $|c^{}_{\pm}|^2$. Without loss of generality, let us assume that $t^{(1)}_{-}=0$, i.e.,
\begin{align}
\label{eq:filtering1}&JJ_{0c}\exp[-i\omega]=X_{c}J_{0b}.
\end{align}
Recalling that the hopping amplitudes are real,\cite{Comment2} the spin-filtering conditions are
\begin{align}
\label{eq:filtering2}&\frac{JJ_{0c}}{J_{0b}}=|X_{c}|=\sqrt{y^{2}_{c}+2y_{c}J_{0}\cos k_{0}+J^{2}_{0}}\ ,\nonumber\\
&\tan\omega=\frac{J_{0}\sin k_{0}}{y_{c}+J_{0}\cos k_{0}}\ .
\end{align}
Similar to the spin-filtering conditions obtained in Refs.~\onlinecite{AA11} and~\onlinecite{MS13a} for two-terminal interferometers, Eq.~(\ref{eq:filtering2}) for the three-terminal configuration also contains two conditions for the full filtering. The first relates the hopping amplitudes within the two different paths. This condition involves the electron energy $E$ and the various hopping amplitudes and on-site energies, and is independent of the SOI. In the linear-response regime, and at low temperatures, all the transport electrons have practically the same energy, equal to the average chemical potential of the leads. The second condition determines the SOI phase $\omega$ acquired upon completing a full turn (from site 0 back to site 0). When these two conditions are fulfilled the spin of the transmitted electrons is fully polarized, $\ket{\chi^{(1)}_{t}}=\ket{\hat{\bold{n}}_{1}}=U_{0b}\ket{\hat{\bold{n}}}$.

A full spin filtering in lead 2 necessitates in addition that either $t^{(2)}_{+}=0$ or $t^{(2)}_{-}=0$. When $t^{(2)}_{+}=0$, the spin-filtering conditions for lead 2 coincide with Eqs.~(\ref{eq:filtering1}) and~(\ref{eq:filtering2}), with the replacements $b\leftrightarrow c$. From the equations for the SOI phase $\omega$ [i.e., the second of Eqs.~(\ref{eq:filtering2}) and its equivalent for lead 2] one concludes that $y_{b}=y_{c}=y$. The first of Eqs.~(\ref{eq:filtering2}) then implies that  $J_{0b}=\pm J_{0c}$. The electrons transmitted into the two output leads are thus fully polarized simultaneously, with their spinors being $\ket{\chi^{(1)}_{t}}=\ket{\hat{\bold{n}}_{1}}=U_{0b}\ket{\hat{\bold{n}}}$ and $\ket{\chi^{(2)}_{t}}=\ket{-\hat{\bold{n}}_{2}}=U_{0c}\ket{-\hat{\bold{n}}}$, provided that
\begin{align}
\label{eq:filtering4}&y_{b}=y_{c}=y\ ,\ \ \ %\nonumber\\
J_{0b}=\pm J_{0c}\ ,\nonumber\\
&J=\pm \sqrt{y^{2}+2y J_0\cos k_0+J^{2}_{0}}\ ,\nonumber\\
&\tan\omega=\frac{J_{0}\sin k_0}{y+J_0\cos k_0}\ .
\end{align}
The first two conditions are trivially fulfilled for a symmetric interferometer, i.e, when the various parameters in the two arms are identical. Then the conditions for full filtering are given by the last two equations. This symmetric case was explored in a ring interferometer in Ref.~\onlinecite{FP06}. However, the conditions~(\ref{eq:filtering4}) can be fulfilled also when the two arms of the interferometer are not precisely  identical, since the condition $y_{b}=y_{c}$ does not require the arms $bd$ and $cd$ to be identical. Specifically, this condition holds for $\varepsilon^{}_{b}\neq\varepsilon^{}_{c}$, provided that $J^2_{cd}-J^2_{bd}=(\varepsilon^{}_{b}-\varepsilon^{}_{c})(E-\varepsilon^{}_d)$ [see the first of Eqs.~(\ref{eq:3ter parameters 1})]. Also, the electrons in the two outgoing channels are fully polarized under the asymmetric condition $J_{0b}=-J_{0c}$.

Similarly, when $t^{(2)}_{-}=0$, the spin-filtering conditions for lead 2 can be written in the same form as Eq.~(\ref{eq:filtering2}) with the replacements $b\leftrightarrow c$ and $\omega\rightarrow -\omega$, and full polarization in both output leads is achieved for disparate interferometer arms. For example, the equations for the SOI phase $\omega$ then imply that $y_{b}+y_{c}=-2J_0\cos k_0$. For appropriate parameters, the spinors of the electrons transmitted into the two output leads in this case are $\ket{\chi^{(1)}_{t}}=\ket{\hat{\bold{n}}_{1}}=U_{0b}\ket{\hat{\bold{n}}}$ and $\ket{\chi^{(2)}_{t}}=\ket{\hat{\bold{n}}_{2}}=U_{0c}\ket{\hat{\bold{n}}}$.

Our analysis suggests that spin filtering by a three-terminal interferometer may also be obtained for
a partially asymmetric interferometer, and thus generalizes the results of Ref.~\onlinecite{FP06}. While tuning of two parameters is sufficient for a full filtering in the two output leads of a symmetric  interferometer, the (probably more realistic) asymmetric one requires additional tuning of $J_{0b}$ or $J_{0c}$, and of either of $\varepsilon_{b}$, $\varepsilon_{c}$, $J_{bd}$, or $J_{cd}$.

In general, when the spin-filtering conditions are not fulfilled exactly, the transmitted electrons in each output lead  are only partially polarized. The spin polarization in the $n$th output lead along $\hat{\bold{n}}_{n}$ is readily calculated by substituting Eqs.~(\ref{eq:transmission eigenvalues}) into Eq.~(\ref{eq:polarization1}),
\begin{align}
\label{eq:polarization3}&P^{(1)}_{\hat{\bold{n}}_{1}}=-\frac{2JJ_{0b}J_{0c}|X_{c}|\sin\omega\sin\delta_{c}}{J^{2}J^{2}_{0c}+J^{2}_{0b}|X_{c}|^{2}-2JJ_{0b}J_{0c}|X_{c}|\cos\omega\cos\delta_{c}}\ ,\nonumber\\
&P^{(2)}_{\hat{\bold{n}}_{2}}=\frac{2JJ_{0b}J_{0c}|X_{b}|\sin\omega\sin\delta_{b}}{J^{2}J^{2}_{0b}+J^{2}_{0c}|X_{b}|^{2}-2JJ_{0b}J_{0c}|X_{b}|\cos\omega\cos\delta_{b}}\ ,
\end{align}
where the variables $X_{b}$ and $X_{c}$ [see Eqs.~(\ref{eq:filtering1}) and~(\ref{eq:filtering2})]
are presented in the form $|X_{b}|\exp[i\delta_{b}]$ and $|X_{c}|\exp[i\delta_{c}]$, respectively. The  polarizations (\ref{eq:polarization3}) achieved in a symmetric interferometer are equal in magnitude and of opposite signs.\cite{comment3} However, as $\hat{\bold{n}}_{1}\neq \hat{\bold{n}}_{2}$, the total polarization~(\ref{eq:polarization2}) does not vanish even in the symmetric case.

\begin{figure}[ht!]
	\begin{center}
		\includegraphics[width=0.52\textwidth,height=0.23\textheight]{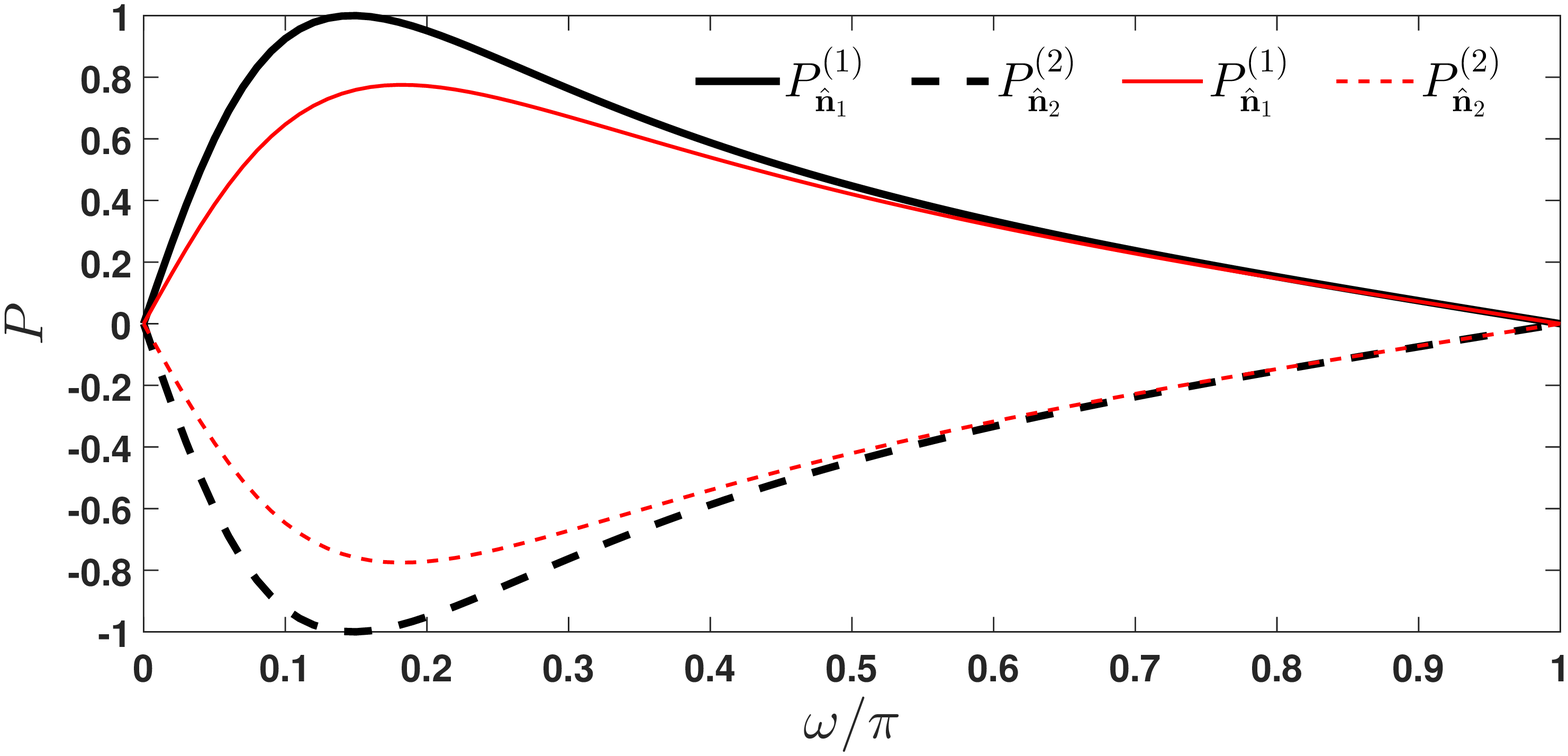}
	\end{center}
	\caption{\label{fig:polarization_interferometer}(Color online) Spin polarization in the first (solid) and second (dashed) output leads, along the directions $\hat{\bold{n}}_{1}$ and $\hat{\bold{n}}_{2}$, respectively, as a function of $\omega$ at the center of the energy band in the leads, $E=0$. For the black (thick) curves $y_{b}=y_{c}=y=2J_{0}$, $J_{0b}=J_{0c}$, and $J=\sqrt{y^{2}+J^{2}_{0}}=\sqrt{5}J_{0}$. Full spin polarization is obtained for $\omega=\arctan(J_{0}/y)\approx 0.15\pi$. For the red (thin) curves $J_{0b}=0.7J_{0c}$ and only a partial spin polarization is obtained.}
\end{figure}

The above results have been obtained from a tight-binding model, with 1D leads and 1D arms of the interferometer. In realistic situations the various wires are not 1D, and therefore our detailed predictions may not apply at arbitrary energies (especially near the van Hove singularities). However, they are expected to be qualitatively valid when the transport in the wire is through a single channel, and for energies near the band center, $E=0$ or $k_{0}=\pi/2$, where the density of states depends only weakly on the energy. Figure~\ref{fig:polarization_interferometer} illustrates the polarization along the direction $\hat{\bold{n}}_{n}$ in each of the output leads as a function of $\omega$, for $E=0$. For the black (thick) curves the first three conditions of Eqs.~(\ref{eq:filtering4}) are satisfied, and thus full polarization is obtained in both leads when the fourth of Eqs.~(\ref{eq:filtering4}) is satisfied, i.e.\ when $\tan\omega=J_{0}/y$. The red (thin) curves correspond to $J^{}_{0b}\neq J^{}_{0c}$, so that the spin polarization is only partial. 
%
%%%%%%%%%%%%%%%%%%%%%%%%%%%%%%%%%%%%%%%%%%%%%%%%%%%%%%%%%%%%
\section{Three-terminal  helical molecule} \label{Sec4}
\begin{figure}[htp]
\includegraphics[width=4cm]{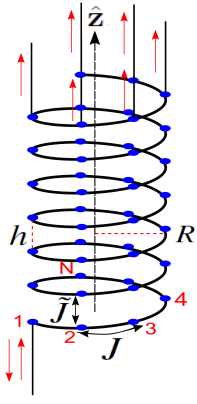}
	\caption{\label{fig:Model}(Color online) Illustration of the model of a single helical molecule, of radius $R$ and pitch $h$. In the tight-binding picture, electrons hop between adjacent sites along the helix (hopping amplitude $J$) and along the vertical direction $\hat{\bf z}$, to the $N$th neighbor (hopping amplitude $\tilde{J}$). Spin-orbit interaction is assumed to act only between nearest neighbors along the helix. The molecule is connected to a single lead at one edge and to $N$ leads at the sites of the last ($M$th) unit cell.}
\end{figure}

This section is devoted to an analysis of spin-dependent transport through a single helical molecule  placed in a multi-terminal configuration, as shown in Fig.~\ref{fig:Model}. The molecule is modeled within the tight-binding formalism: the hopping amplitude between nearest-neighbor sites along the helix is $J$ and that between sites along the axial direction $\hat{\bf z}$, whose distance is $h$ ($h$ being the pitch of the helix, see Fig.~\ref{fig:Model}), is $\tilde{J}$. The on-site energies are set to zero. It is assumed that the SOI is effective only between nearest neighbors along the helix (and not between the axial bonds). An analytical solution can then be obtained for a unit cell with $N$ sites along the helix, assuming an arbitrary set of unitary matrices
\begin{align}
U_{n}=\exp[i\mathbf{K}_{n,n+1}\cdot\boldsymbol{\sigma}]\ ,\ \    n=1\ , \ldots\ ,N\ ,
\label{Un}
\end{align}
which describe the spin precession of an electron moving between the nearest-neighbor sites $n$ and $n+1$ in each unit cell. This description of the helical molecule is similar to that investigated in previous works~\cite{GAM12,GAM14a,WHN16,MS16} in a two-terminal geometry, where TRS or unitraity is effectively broken, as discussed in Sec.~\ref{Sec1}. Here we solve for the unitary scattering matrix of this model when it includes $M$ unit cells; the molecule is attached to a single lead at one of its edges, and to $N$ leads at the other edge, by one-dimensional chains on which the hopping amplitudes are all $J_{0}$ and the on-site energies are zero.

As in Sec.~\ref{Sec3}, we assume that all leads save one (``output leads") have the same chemical potential, which is lower than that in the source reservoir attached to the lead singled out (``input lead"). The results are demonstrated explicitly on a molecule with two output leads, i.e., for a three-terminal setup.

The solution of the scattering problem, which yields the scattering matrix, is carried out for a periodic structure which contains $M$ unit cells; the Schr\"{o}dinger equation is solved for a scattering state of the form~(\ref{eq:scattering state 3ter}) for the spinors $\ket{\psi^{}_{m,n}}$ inside the molecule, with $1\leq m\leq M$ and $1\leq n\leq N$. The first index indicates the unit cell and the second the site within that cell. The continuity conditions on the scattering state are then
\begin{align}
\label{eq:continuity}&\ket{\chi^{}_{in}}+r\ket{\chi^{}_{r}}=\ket{\psi^{}_{1,1}}\ ,\nonumber\\
&t^{(n)}\ket{\chi^{(n)}_{t}}=\ket{\psi^{}_{M,n}}\ , \quad 1\leq n\leq N\ .
\end{align}
The dispersion relation relating the energy of the electron to its wave vector is\cite{MS16}
\begin{align}
\label{eq:dispersion relations}&E_{p,\sigma}(k)=-2\tilde{J}\cos k-2J\cos\left(\frac{k+2\pi p-\sigma\theta}{N}\right)\ .
\end{align}
Here, $k$ is the wave vector in units of $(N\ell)^{-1}$, where $\ell$ is the distance between nearest neighbors along the helix [see Eq.~(\ref{el})], and $p=1,\ldots,N$ is the band index.
The spin index $\sigma=\pm 1$ is the eigenvalue corresponding to the eigenspinor $\ket{\pm\hat{\bold{n}}}$ of the spin projection along $\hat{\bold{n}}$. This direction is defined by the unitary matrix
\begin{align}
\label{eq:V}\mathcal{U}=U_{N}U_{N-1}\times\ldots\times U_{1}\equiv \exp[i\theta\,\hat{\bold{n}}\cdot\boldsymbol\sigma]\ .
\end{align}
The spin precession angle per one turn of the helix, $\theta$, is equivalent to the SOI phase $\omega$ introduced in Sec.~\ref{Sec3}. For a given energy $E$ and spin $\sigma$, the dispersion relation~(\ref{eq:dispersion relations}) yields $2N$ solutions for the variable $y=\left(k+2\pi p\right)/N$, and the spinor inside the molecule is a linear combination of these solutions,~\cite{MS16}
\begin{align}
\label{eq:psimn}\ket{\psi_{m,n}}&=U^{\dag}_{n}\ldots U^{\dag}_{N}\nonumber\\
&\times\sum_{\sigma=\pm1}
\sum_{j=1}^{2N}A^{\sigma}_{j} e^{i\left[y^{\sigma}_{j}\left(mN+n\right)-\sigma\theta n/N\right]}\ket{\sigma\hat{\bold{n}}}\ .
\end{align}
The amplitudes $A^{\sigma}_{j}$ are determined by the Schr\"{o}dinger  equations for the spinors in the first ($m=1$) and the last ($m=M$) unit cells,
%\begin{widetext}
	\begin{align}
	\label{eq:m=1}&\left(E-y_{0}\right)\ket{\psi^{}_{1,1}}=2iJ_{0}\sin k_{0}\ket{\chi^{}_{in}}-JU^{\dag}_{1}\ket{\psi^{}_{1,2}}-\tilde{J}\ket{\psi^{}_{2,1}}\!\ ,\nonumber\\
	%E\ket{\psi^{}_{1,n}}&=-JU_{n-1}\ket{\psi^{}_{1,n-1}}-JU^{\dag}_{n}\ket{\psi^{}_{1,n+1}}\nonumber\\
	%&-\tilde{J}\ket{\psi^{}_{2,n}}\!\ , %\quad 2\leq n\leq N-1\ ,
	%\nonumber\\
	&E\ket{\psi^{}_{1,N}}=-JU_{N-1}\ket{\psi^{}_{1,N-1}}-JU^{\dag}_{N}\ket{\psi^{}_{2,1}}-\tilde{J}\ket{\psi^{}_{2,N}}\!\ ,\nonumber\\
	&E\ket{\psi^{}_{1,n}}=-JU_{n-1}\ket{\psi^{}_{1,n-1}}-JU^{\dag}_{n}\ket{\psi^{}_{1,n+1}}-\tilde{J}\ket{\psi^{}_{2,n}}\!\ ,% \quad 2\leq n\leq N-1\ ,
	\end{align}
	 and	
	\begin{align}
	\label{eq:m=M}\left(E-y_{0}\right)\ket{\psi^{}_{M,1}}&=-JU^{\dag}_{1}\ket{\psi^{}_{M,2}}-JU_{N}\ket{\psi^{}_{M-1,N}}\nonumber\\
	&-\tilde{J}\ket{\psi^{}_{M-1,1}}\!\ ,\nonumber\\
	\left(E-y_{0}\right)\ket{\psi^{}_{M,n}}&=-JU_{n-1}\ket{\psi^{}_{M,n-1}}-JU^{\dag}_{n}\ket{\psi^{}_{M,n+1}}\nonumber\\
	&-\tilde{J}\ket{\psi^{}_{M-1,n}}\!\ ,
	\nonumber\\% \quad 2\leq n\leq N-1\ ,\nonumber\\
	\left(E-y_{0}\right)\ket{\psi^{}_{M,N}}&=-JU_{N-1}\ket{\psi^{}_{M,N-1}}-\tilde{J}\ket{\psi^{}_{M-1,N}}\!\ ,
	\end{align}
%\end{widetext}
where $2\leq n\leq N-1$ and $y_{0}=-J_{0}\exp[ik_{0}]$. When $\ket{\chi^{}_{in}}=\ket{\hat{\bold{n}}}$, then $A^{-}_{j}=0$, and all the spinors $\ket{\psi^{}_{m,n}}$ have only $\sigma=1$. Similarly, when
$\ket{\chi^{}_{in}}=\ket{-\hat{\bold{n}}}$, then $A^{+}_{j}=0$, and all the spinors $\ket{\psi^{}_{m,n}}$ have only $\sigma=-1$. The remaining amplitudes are found by solving the $2N$ linear equations
\begin{widetext}
	\begin{align}
	\label{eq:amplitudes}&\sum^{2N}_{j=1}\left[E-y_{0}+Je^{i\left(y^{\sigma}_{j}-\sigma\theta/N\right)}+\tilde{J}e^{iy^{\sigma}_{j}N}\right]e^{i\left[y^{\sigma}_{j}\left(N+1\right)-\sigma\theta/N\right]}A^{\sigma}_{j}=2iJ_{0}\sin k_{0}e^{i\sigma\theta}\ ,\nonumber\\
	&\sum^{2N}_{j=1}\left[E+2J\cos\left(y^{\sigma}_{j}-\sigma\theta/N\right)+\tilde{J}e^{iy^{\sigma}_{j}N}\right]e^{iy^{\sigma}_{j}\left(N+n\right)}A^{\sigma}_{j}=0\ , \quad 2\leq n\leq N\ ,\nonumber\\
	&\sum^{2N}_{j=1}\left[E-y_{0}+2J\cos\left(y^{\sigma}_{j}-\sigma\theta/N\right)+\tilde{J}e^{-iy^{\sigma}_{j}N}\right]e^{iy^{\sigma}_{j}\left(MN+n\right)}A^{\sigma}_{j}=0\ , \quad 1\leq n\leq N-1\ ,\nonumber\\
	&\sum^{2N}_{j=1}\left[E-y_{0}+Je^{-i\left(y^{\sigma}_{j}-\sigma\theta/N\right)}+\tilde{J}e^{-iy^{\sigma}_{j}N}\right]e^{iy^{\sigma}_{j}N\left(M+1\right)}A^{\sigma}_{j}=0\ .
	\end{align}
\end{widetext}
As seen from  Eq.~(\ref{eq:continuity}), the reflected electron is  polarized along $\sigma\hat{\bold{n}}$ for $\ket{\chi_{in}}=\ket{\sigma\hat{\bold{n}}}$, whereas for the transmitted electron in the $n$th output lead
\begin{align}
\label{eq:helix transmitted}t^{(n)}\ket{\chi^{(n)}_{t}}&=U^{\dag}_{n}\ldots U^{\dag}_{N}\nonumber\\
&\times\sum^{2N}_{j=1}A^{\sigma}_{j}e^{i\left[y^{\sigma}_{j}\left(MN+n\right)-\sigma\theta n/N\right]}\ket{\sigma\hat{\bold{n}}}\!\ ,
\end{align}
and therefore
\begin{align}
\label{eq:helix transmitted spinors}&\ket{\chi^{(n)}_{t}}=\ket{\sigma\hat{\bold{n}}_{n}}\equiv U^{\dag}_{n}\ldots U^{\dag}_{N}\ket{\sigma\hat{\bold{n}}}\!\ .
\end{align}
The solution for the $2\times2$ reflection and transmission amplitude matrices $\mathcal{R}$ and $\mathcal{T}_{n}$, defined in Eq.~(\ref{eq:R and T matrices 1}), has the form~(\ref{eq:R and T matrices 2}) with the reflection and transmission amplitudes
\begin{align}
\label{eq:reflection transmission}&r=-1+e^{-i\theta\left(1+1/N\right)}\sum^{2N}_{j=1}A^{+}_{j}e^{iy^{+}_{j}\left(N+1\right)}\ ,\nonumber\\
&t^{(n)}_{\pm}=e^{\mp i\theta n/N}\sum^{2N}_{j=1}A^{\pm}_{j}e^{iy^{\pm}_{j}\left(MN+n\right)}\ ,
\ \ 1\leq n\leq N\ .
\end{align}
The solution of the scattering problem at a given energy $E$ requires finding the $2N$ roots $y^{\sigma}_{j}$ of a polynomial of degree $2N$, and then solving the system of $2N$ equations~(\ref{eq:amplitudes}) for the amplitudes $A^{\sigma}_{j}$. Therefore, in contrast to the diamond interferometer presented in Sec.~\ref{Sec3}, no simple analytic expression for the transmission amplitudes $t^{(n)}_{\pm}$ can be obtained for our model of the molecule.

The model presented above is constructed for a molecule with $N$ sites in the unit cell which repeat themselves, and with a coupling between neighboring unit cells. However, up to this point the hopping amplitudes between nearest-neighbor sites within the unit cell, described by the unitary matrices $U_{n}$ [Eq.~(\ref{Un})], are arbitrary, allowing e.g., for different atoms (or side groups) within the cell, as might be expected for realistic organic molecules.

To demonstrate the results, we apply the model to periodic helical molecules, where all the nearest-neighbor bonds are equivalent except for a rotation around the helix axis and a shift by the helix pitch. We also assume that the vector $\mathbf{K}_{n,n+1}$ in Eq.~(\ref{eq:SOI matrices}) is of the form~(\ref{eq:K vectors 2}). For a helical molecule of radius $R$ and pitch $h$, the vector $\mathbf{d}_{n,n+1}$ is
\begin{align}
\label{eq:vector d}&\mathbf{d}_{n,n+1}=2R\sin\left(0.5\Delta\varphi\right)\left(-s_{n}\hat{\mathbf{x}}+c_{n}\hat{\mathbf{y}}\right)+\frac{h}{N}%\left(h/N\right)
\hat{\mathbf{z}}\ ,
\end{align}
where $s_{n}=\sin\left[\left(n+0.5\right)\Delta\varphi\right]$, $c_{n}=\cos\left[\left(n+0.5\right)\Delta\varphi\right]$ and $\Delta\varphi=\pm 2\pi/N$ is the twist angle between nearest-neighbor sites, with the plus (minus) sign corresponding to right-handed (left-handed) chirality. Assuming also that the SOI is induced by the confinement of the electron to the cylinder which contains the helix, the electric field is taken as constant in the radial direction,
\begin{align}
\label{eq:Electric field}&\mathbf{E}_{n,n+1}=E_{0}\left(c_{n}\hat{\mathbf{x}}+s_{n}\hat{\mathbf{y}}\right)\ .
\end{align}
For this helical geometry, %we can calculate
the spin rotation angle $\theta$ and the directions $\hat{\bold{n}}_{n}$ are given  in terms of $R$, $h$, the length of each bond
\begin{align}
\ell=\sqrt{\left(h/N\right)^{2}+\left[2R\sin\left(0.5\Delta\varphi\right)\right]^{2}}\ ,
\label{el}
\end{align}
the spin precession angle per bond $\tilde{\lambda}=\lambda E_{0}\ell$, and the chirality of the molecule (specified by the sign of $\Delta\varphi$). In particular, in the three-terminal configuration ($N=2$),
the matrix~(\ref{eq:V}) yields that the rotation angle $\theta$ is
\begin{align}
\label{eq:theta and n}&\cos\theta=\cos(2\tilde{\lambda})+\frac{h^{2}}{2\ell^{2}}\sin^{2}\tilde{\lambda}\ ,
\end{align}
and the spin projection direction $\hat{\bold{n}}$ [see Eq.~(\ref{eq:V})] lies in the YZ plane,
\begin{align}
%n_{x}=0\ ,
n_{y}\sin\theta=\frac{2hR}{\ell^{2}}\sin^{2}\tilde{\lambda}\ ,\ %\nonumber\\
n_{z}\sin\theta=\mp\frac{2R}{\ell}\sin(2\tilde{\lambda})\ .
\label{nvec}
\end{align}
For these values, Eq.~(\ref{eq:helix transmitted spinors}) yields $\hat{\bold{n}}_{1}=\hat{\bold{n}}$ and $\hat{\bold{n}}_{2}=\left(\sin\alpha\cos\beta,\sin\alpha\sin\beta,\cos\alpha\right)$,  where
\begin{align}
\label{eq:n_2}&\cos\alpha=\pm\frac{h}{2\ell}\sin(2\tilde{\lambda})n_{y}+\left(1-\frac{h^{2}}{2\ell^{2}}\sin^{2}\tilde{\lambda}\right)n_{z}\ ,\nonumber\\
&\tan\beta=\frac{\frac{1}{2}\cos(2\tilde{\lambda})n_{y}\mp\frac{h}{4\ell}\sin(2\tilde{\lambda})n_{z}}{\mp\frac{R}{\ell}\sin(2\tilde{\lambda})n_{y}+\frac{h}{2\ell}\left(\frac{2R}{\ell}\sin^{2}\tilde{\lambda}+\frac{1}{2}\sin(2\tilde{\lambda})\right)n_{z}}\ .
\end{align}

\begin{figure}[ht!]
	\begin{center}
		\includegraphics[width=0.52\textwidth,height=0.23\textheight]{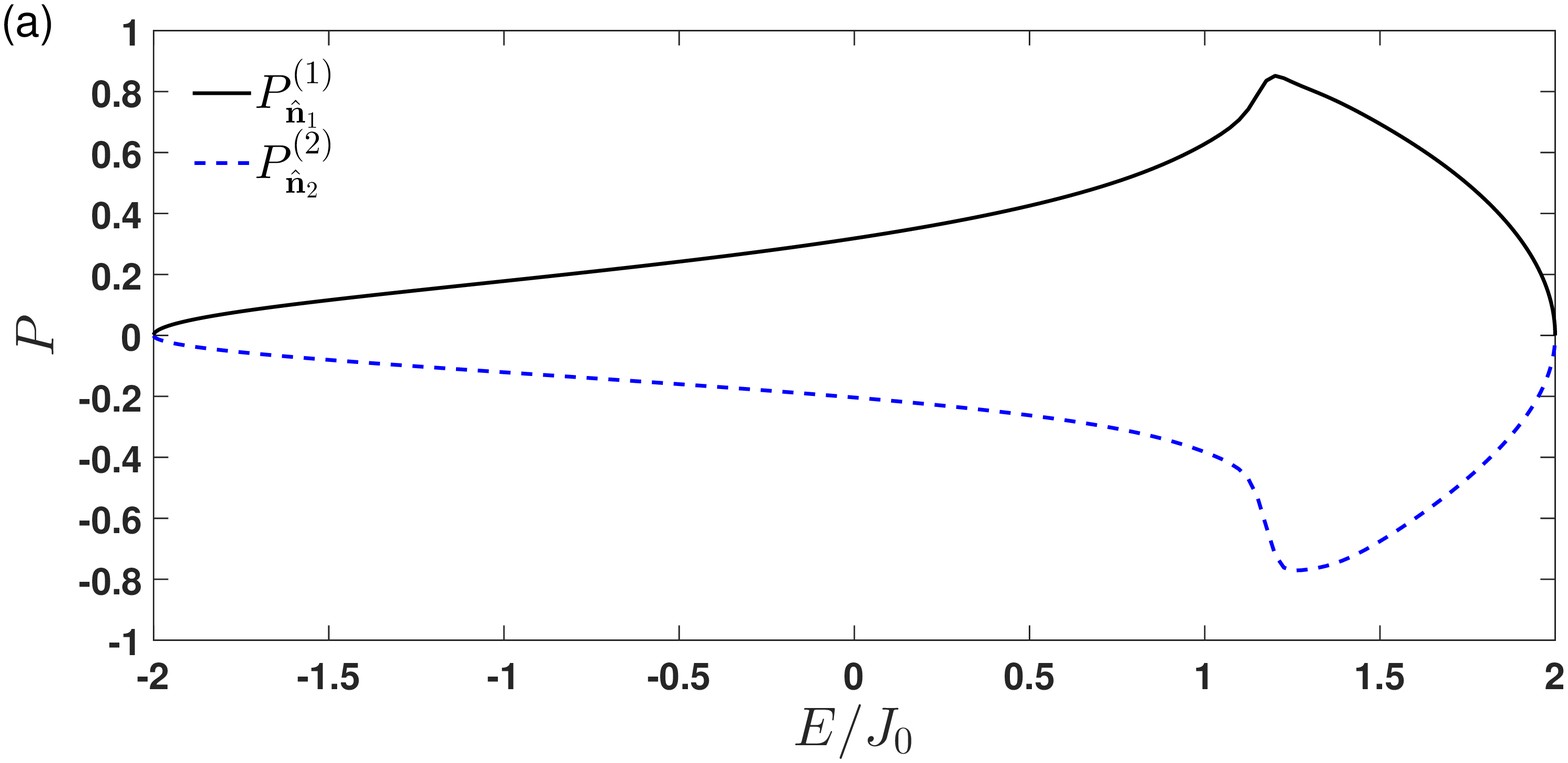}
		\includegraphics[width=0.52\textwidth,height=0.23\textheight]{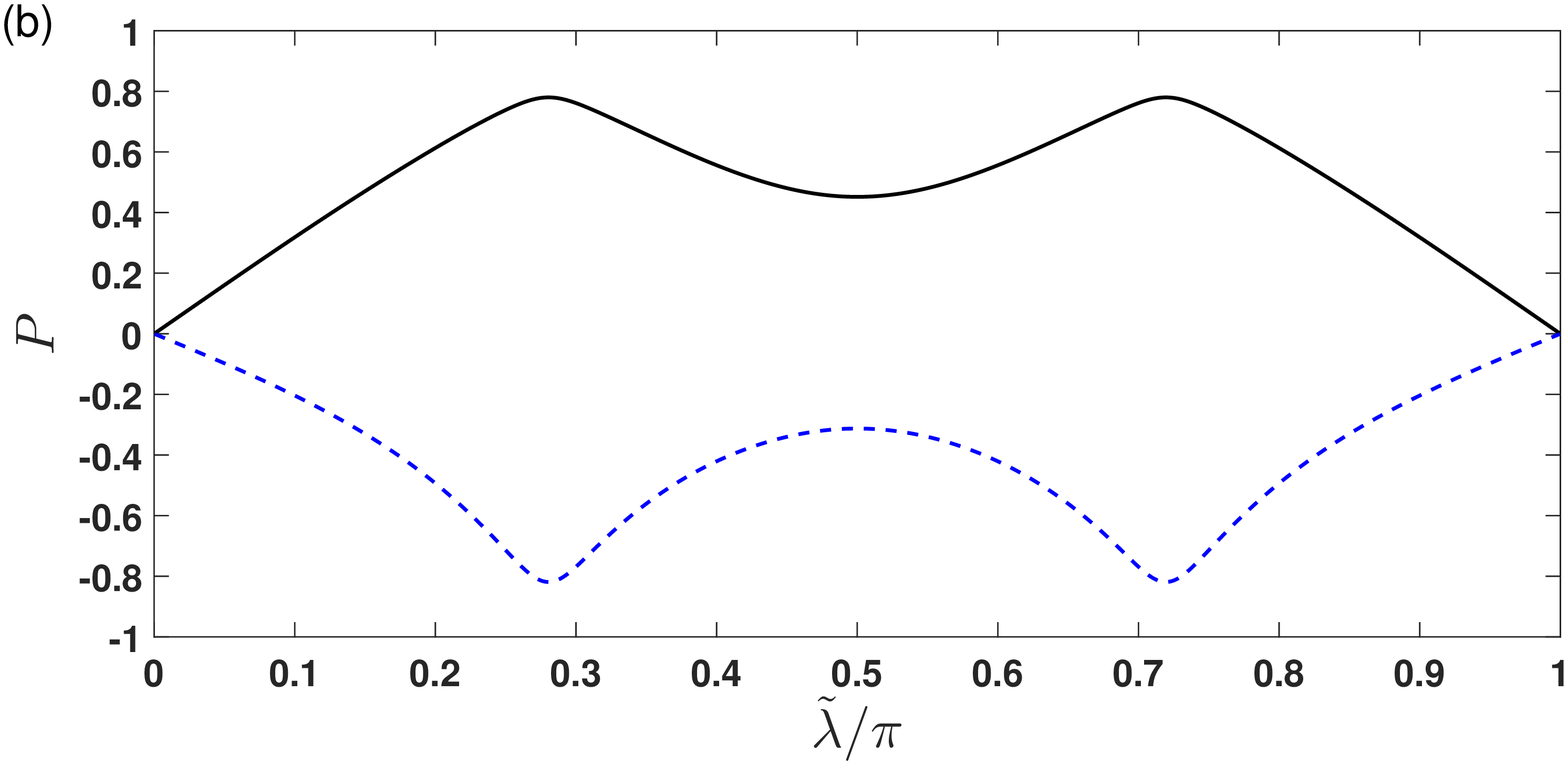}
	\end{center}
	\caption{\label{fig:polarization1}(Color online) Spin polarization in the first (solid black) and second (dashed blue) output leads, along the directions $\hat{\bold{n}}_{1}$ and $\hat{\bold{n}}_{2}$, respectively, in a molecule of $M=6$ unit cells (the other parameters are $J=1.2J^{}_{0}$ and $\tilde{J}=0.5J^{}_{0}$). (a) As a function of energy (in units of $J^{}_{0}$) with $\tilde{\lambda}=0.1\pi$; (b) as a function of $\tilde{\lambda}$ at the center of the energy band in the leads, $E=0$.}
\end{figure}

The polarization along the direction $\hat{\bold{n}}_{n}$ in each of the output leads,
given by Eq.~(\ref{eq:polarization1}), is depicted in Fig.~\ref{fig:polarization1}(a) as a function of the electron energy $E$, and in Fig.~\ref{fig:polarization1}(b) as a function of the spin precession angle per bond $\tilde{\lambda}$. This is the direction along which the polarization in each output lead is maximal.~\cite{MS16}
\begin{figure}[ht!]
	\begin{center}
		\includegraphics[width=0.52\textwidth,height=0.23\textheight]{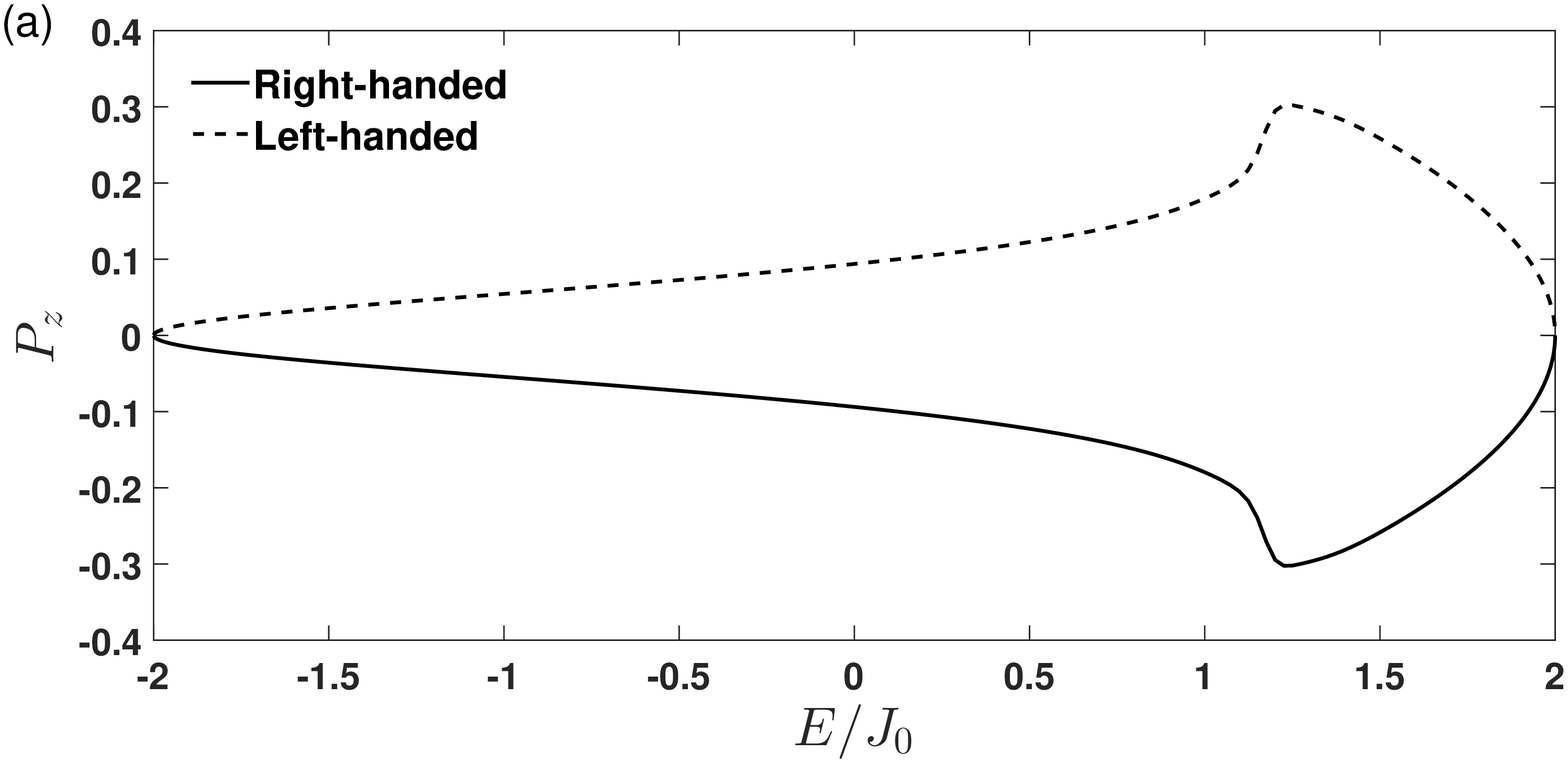}
		\includegraphics[width=0.52\textwidth,height=0.23\textheight]{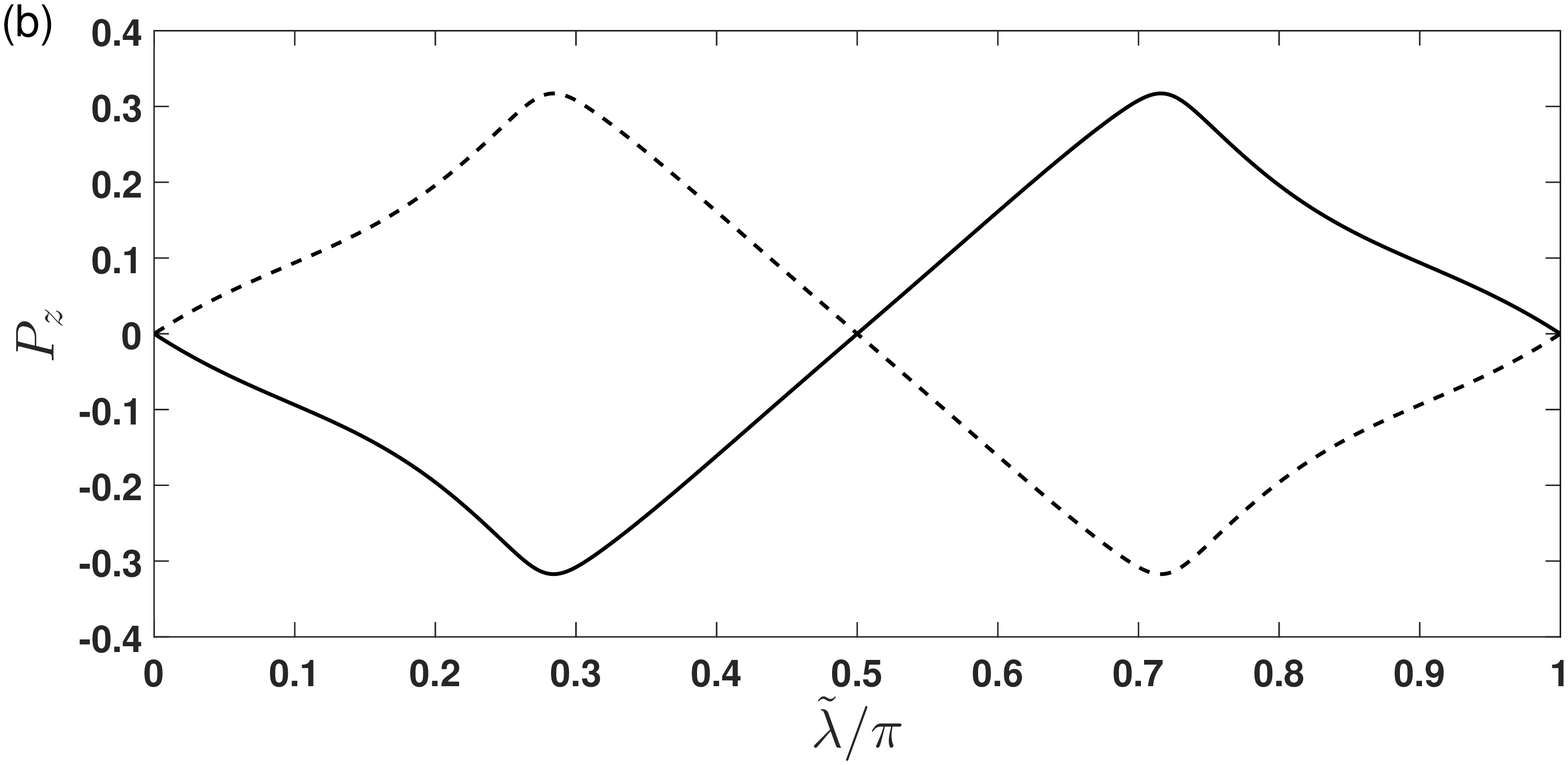}
	\end{center}
	\caption{\label{fig:polarization2} Total spin polarization along the $z$-axis in a right-handed (solid) and left-handed (dashed) molecule of $M=6$ unit cells with $h/\ell=0.8$ (the other parameters are as in Fig.~\ref{fig:polarization1}). (a) As a function of energy (in units of $J^{}_{0}$) with $\tilde{\lambda}=0.1\pi$; (b) as a function of $\tilde{\lambda}$ at the center of the energy band in the leads, $E=0$.}
\end{figure}

Assuming that all the outgoing electrons are eventually mixed incoherently, the total spin polarization along the $\hat{\bf z}$ direction is calculated from Eq.~(\ref{eq:polarization2}). It is displayed in Fig.~\ref{fig:polarization2}(a) as a function of the energy, and in Fig.~\ref{fig:polarization2}(b) as a function of the spin precession angle per bond $\tilde{\lambda}$. As follows from Eqs.~(\ref{nvec}) and~(\ref{eq:n_2}), this polarization changes sign as the chirality of the molecule is reversed. This feature is in agreement with the experimental observation.~\cite{NR15}
%%%%%%%%%%%%%%%%%%%%%%%%%%%%%%%%%%%%%%%%%%%%%%%%%%%%%%%%%%%%
\section{Summary and discussion} \label{Summary}
Spin-dependent transport through coherent multi-terminal mesoscopic systems in which the transport electrons are subjected to spin-orbit interactions are analyzed. Explicit results are demonstrated for three-terminal configurations. In contrast to the two-terminal configuration, in a multi-terminal setup the transmitted electrons in each lead are in general spin-polarized even in the absence of magnetic fields. Specifically, it is found that a simple three-terminal two-path interferometer can act as an all-electrical spin beam-splitter. This device splits the incoming electrons into two fully spin-polarized beams, with tunable spin directions, as first discussed in Ref.~\onlinecite{FP06} for a symmetric ring interferometer. We have shown that such an interferometer can serve as a perfect spin filter even in the more realistic case where its two arms are not completely identical. The conditions for full spin-filtering in the general asymmetric case and the resulting spin polarization in the two output leads are derived analytically. The simple model for the single interferometer was used to demonstrate analytically that in a three-terminal configuration the total polarization of all the transmitted electrons does not vanish in general.

In the second part of the paper we have argued that the nonzero total polarization in a three-terminal configuration may be relevant for the explanation of the recently observed spin selectivity in chiral organic molecules. We have solved a minimal model for a chiral helical molecule in a three-terminal configuration, in which electrons can exit the molecule into two terminals. This model is similar to the one of Refs.~\onlinecite{GAM12}, \onlinecite{GAM14a} and~\onlinecite{MS16}, but does not involve any kind of fictitious probes that are introduced to break time-reversal symmetry or unitarity. It suggests that a significant spin polarization can be obtained in a multi-terminal configuration assuming a completely coherent and unitary electron transport.

\begin{acknowledgments}
AA and OEW acknowledge the hospitality of the Centro Atomico Bariloche and Instituto Balseiro in Bariloche, Argentina, where part of this work was accomplished. This work was supported by the Israeli Science Foundation (ISF) and by the infrastructure program of Israel Ministry of Science and Technology under contract 3-11173.
\end{acknowledgments}


\begin{references}
\bibitem{Comment1} All the wave vectors are in units of the inverse of the corresponding lattice constant.
\bibitem{comment3} This is correct also for an asymmetric interferometer, when the first and either the second or the third of Eqs.~(\ref{eq:filtering4}) are satisfied.
\end{references}
\end{document}